# Exploring an Application of Virtual Reality for Early Detection of Dementia

Yiming Zhong, Yuan Tian, Mira Park, Soonja Yeom

Centre of Digital Technologies Education Research (CODER)

ICT, School of Technology, Environments & Design

University of Tasmania, Australia

*mira.park@utas.edu.au*

## Abstract

Facing the severe global dementia problem, an exploration was conducted adopting the technology of virtual reality (VR). This report lays a technical foundation for further research project "*Early Detection of Dementia Using Testing Tools in VR/AR Environment*", which illustrates the process of developing a VR application using Unity 3D software on Oculus Go. This preliminary exploration is composed of three steps, including 3D virtual scene construction, VR interaction design and monitoring. The exploration was recorded to provide basic technical guidance and detailed method for subsequent research.

Keywords: dementia, early detection, dementia diagnosis, virtual reality, future directions, Oculus Go, Unity.

## 1. Introduction

The number of people enduring dementia in the world is 46 million nowadays, and it is expected to soar to 131.5 million by 2050 (Hayhurst, 2017). This data shows that dementia has become an issue that cannot be ignored. In regarding to Australia, over 400,000 Australians currently suffer from dementia, and by 2058, the number is estimated to be 1 million (Australia, 2019). With the continuous improvement and development of VR technology, it is expected to play an active role in the field of dementia. This article illustrates the conduction of a preliminary technical exploration for the early detection and diagnosis of dementia. After a brief literature review, a VR prototype was designed and developed with different 3D scenes. In addition, the contribution of this exploration and subsequent work are illustrated in this report.

# 2. Literature Review

Symptoms that occur when the brain is affected by certain diseases or conditions are named dementia (Hayhurst, 2017). The main symptoms of dementia are attention reduction, memory loss, increasing difficulties in planning and performing tasks, and changes of emotion and personality, which will greatly affect the quality of daily life of people suffering from dementia (Hayhurst, 2017).

The main areas where technology can play a positive role in the field of dementia are as follows: "(i) diagnosis, assessment and monitoring, (ii) maintenance of functioning, (iii) leisure and activity, (iv) caregiving and management" (Astell et al., 2019).

Amongst the different utilized technologies, VR is an example, since VR cognitive training has been proven to be a positive approach to improve subjects' attention and memory functions (Gamito et al., 2015). VR is a three-dimensional-environment simulation generated by computer, and with the help of virtual reality equipment, such as VR gloves and head-mounted displays, people can achieve interaction with the simulated environment (Hodge et al. 2018). In addition, VR technology currently has been applied to explore different stages of Alzheimer disease and related cognitive disorders with various interaction types, different methodology types and distinct immersion types (see Figure 1).

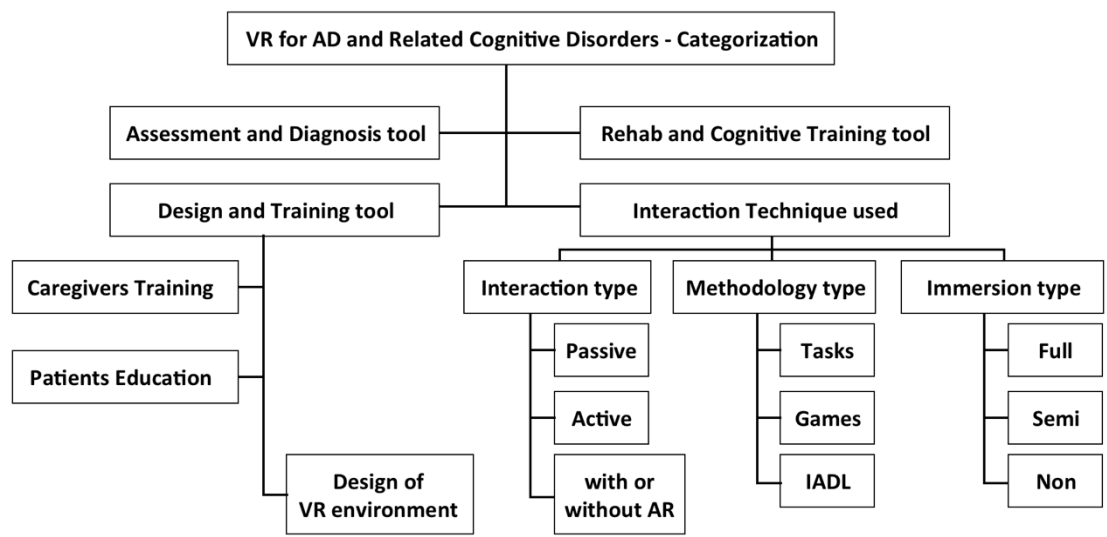

Figure 1   Categorization of VR technology applications for AD. (García-Betances et al., 2015)

Rooted in the succinct review of literature, with the rapid development of VR technology, VR technology is actively applied to several aspects in the field of dementia, including early diagnosis, maintenance of dementia, and training of caregivers and so on. Moreover, it is estimated that VR technology will be useful and helpful in the diagnosis and maintenance of dementia. Because VR technology not only has advantages in creating highly immersive three-dimensional scenes, but also provides rich interactions between human and computers which can fulfill different customized tasks in various scenarios.

# 3. Prototype Design and Implementation

This section consists of prototype design and prototype implementation, which elaborates the process of developing a prototype from selection of hardware and software to the accomplishment of four 3D VR models. Meanwhile, the interaction between the users and the devices and how to monitor the whole interaction process are illustrated.

## 3.1. Prototype Design

During the stage of prototype design, two problems should be solved. One is to select out a VR device. In addition, the other is to choose an appropriate software. In this exploration, Oculus Go and Unity were utilized as the hardware and the main software platform respectively.

### 3.1.1. Hardware Selection

VR technology needs to be implemented through Head-mounted displays (HMDs), and users need to wear VR headsets for head and eye tracking, image or video projection, sound playback, etc. According to a data report released by Strategy Analytics in 2018, the mainstream VR device brands on the market are Sony, Oculus series, HTC, Google, and Samsung (shown in Figure 2).

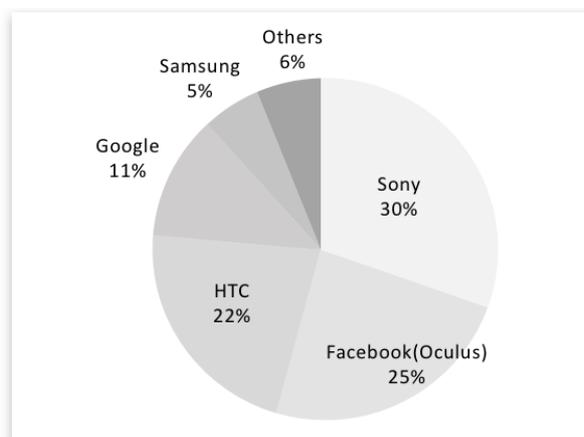

**Figure 2** 2018 VR Headset Revenue Vendor Global Market Share (Strategy Analytics, 2018)

The basic technical specifications of VR equipment include headset type, connection, and resolution, refresh rate (per eye), degree of freedom (DOF), controls, hardware platform, and software platform. On the basis of current mainstream VR headsets, these eight aspects are compared against five mainstream headsets including Sony PlayStation, Oculus Go, Oculus Quest, Oculus Rift 5, and HTC Vive. As can be seen from the comparison below presented in Table 1, Oculus Go has advantages in portability and image display. Although it cannot achieve complex VR interactions such as walking, jumping and crouching, it can meet basic interaction requirements such as moving the head to experience a 360-degree virtual space and using the controller to achieve movement and other interactions in VR space.

Table 1 Comparison of mainstream devices in the VR market (Greenwald, 2019)

|  | Sony PlayStation VR | Oculus Go | Oculus Quest | Oculus Rift 5 | HTC Vive |
|---|---|---|---|---|---|
| Headset Type | Tethered | Standalone | Standalone | Tethered | Tethered |
| Connection | HDMI, USB 2.0 | None | None | DisplayPort, USB 3.0 | HDMI, USB 3.0 |
| Resolution (per eye) | 960 * 1080 | 1280 * 1440 | 1440 * 1600 | 1280 * 1440 | 1080 * 1200 |
| Refresh Rate (Hz) | 120 | 72 | 72 | 80 | 90 |
| Degree-of-Freedom (DOF) | 6DoF | 3DoF | 6DoF | 6DoF | 6DoF |
| Controls | DualShock 4, PlayStation Move | Oculus Go Controller | Oculus Touch | Oculus Touch | HTC Vive Motion Controller |
| Hardware Platform | PlayStation 4 | Android | Android | PC | PC |
| Software Platform | PlayStation 4 | Oculus | Oculus | Oculus | SteamVR |

For further prototype development, a hardware should be selected based on the requirements of the research project. Considering that the main user group of the application is the middle-aged and elderly people who may or already have dementia, the ease of use should be one of the criteria, which is also crucial. Moreover, the controllability and comfort of devices are important, which introduces another criterion Light-weighted standalone device. As for the display quality and interactive functions, it only needs to meet the basic requirements. Hence, a system of hardware criteria was established for this exploration, which is presented in Table 2.

Table 2 Hardware Criteria

| Index | Hardware Criteria |
|---|---|
| 1 | Easy to use, especially for people who are new to VR |
| 2 | Light-weighted standalone device |
| 3 | Normal resolution which can display simple scenes smoothly |
| 4 | Normal tracking which can track the basic position of head |

With the existence of aforementioned comparison between different devices and hardware criteria, Oculus Go was chosen as the hardware to assist further development. Here are some reasons to select it out. Firstly, it (shown in Figure 3) is a standalone headset which means it does not require additional equipment such as a computer and can run independently. Moreover, it is able to achieve 3DoF which is front-back, up-down, and left-right tracking (Oculus.com, 2019). Although the functions of pitch, yaw, and roll are missing compared to 6DoF, many simple VR interactions can be achieved. In addition, as an entry-level VR device, easy operation is the main advantage of Oculus Go. In addition, it is also very friendly to users who has never been exposed to VR devices.

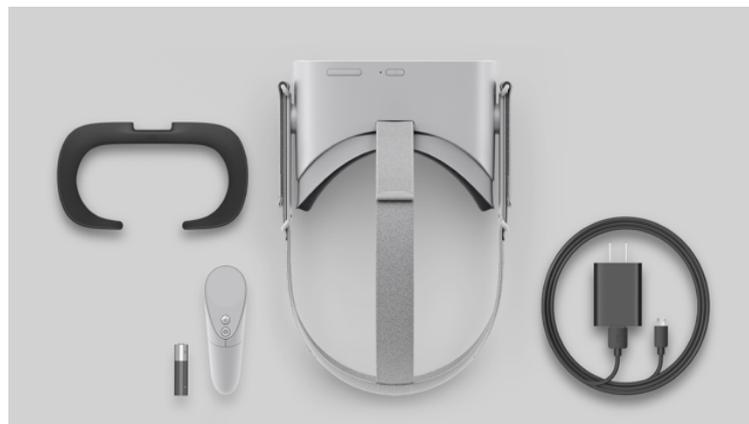

**Figure 3** Oculus Go (www.oculus.com/go)

### 3.1.2. Software Selection

The development of VR technology requires professional software technology and platforms, which is able to adapt to the headset, and provides corresponding resources and plug-ins to support development. It also needs to have the ability to develop 3D models and support programming for interaction. Hence, the criteria to select out an appropriate software was demonstrated in Table 3.

Table 3 Software Criteria

| **Index** | **Software Criteria** |
| --- | --- |
| 1 | Mature development environment |
| 2 | Capable to build 3D models |
| 3 | Able to program basic VR interaction |
| 4 | Competent to collect performance data of users |
| 5 | Available of supporting resources, like plugins |

Unity was selected as the main development platform. Firstly, it is a game development tool that supports multiple platforms and is easy to operate. Moreover, it also has a powerful engine and fully integrated professional application technology (Unity.com, 2019). Furthermore, Unity has become the mainstream VR development software on the market, with its own strong development ecosystem. For instance, it has rich plugins and resource support. Apart from Unity, some other development platforms and software technologies were selected out for this prototype. All of the software utilized in the prototype implementation process and their respective functions were presented in Table 4.

Table 4 Software Utilized in the Prototype Implementation Process

| Name | Function |
| --- | --- |
| Unity | Providing a platform for VR application development, and supporting various VR headsets including Oculus Go |
| OVR Utilities Plugin | Achieving convenient connection between Oculus headsets and Unity |
| ProBuild | Simplifying the processes of model building |
| VReasy | Containing basic VR interaction code, for example, click, trigger animation, pop-up window and controller settings |
| Sketch | Facilitating the graphic and icon design |

Additionally, Figure 4 also shows the key software and hardware platform on which the entire prototype is based. The entire application is primarily built on the Unity 3D platform, and uses the resources of the Unity app store to build models and interactions. After the application is released through the Oculus platform, users can run it directly on Oculus Go.

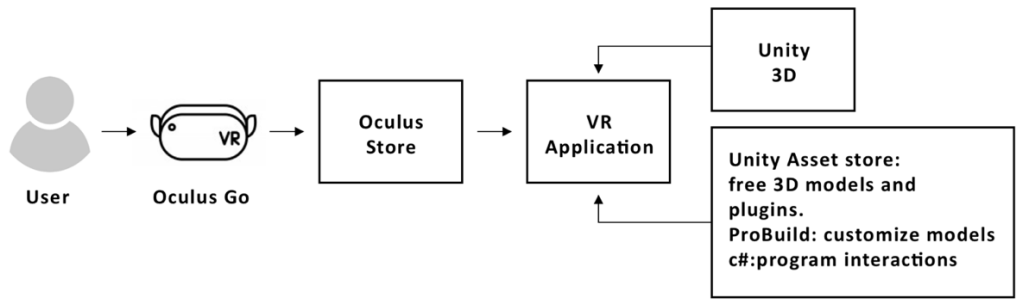

**Figure 4** Key Components of the Prototype

## 3.2. Prototype Implementation

During the stage of prototype implementation, three steps were conducted, which were modelling VR scenarios, interacting with VR equipment and monitoring the interaction between the users and the VR device.

In addition, three-dimensional virtual environment can bring users a sense of immersion. In order to enhance the feeling, the specific model settings are mainly based on daily life scenarios, including common indoor and outdoor scenes. Furthermore, because of the immersive interaction, these scenarios could be utilized to assess the navigational skills of people with dementia. By monitoring and recording the user's behavior in the virtual environments, a meaningful collection of data could be provided for the diagnosis of dementia.

**Step 1. Modelling VR Scenarios**

Some daily scenes were set up in this step, including four different scenarios, which are office rooms (shown in Figure 5), wooden houses (shown in Figure 6), modern interiors (shown in Figure 7) and parks. The model is mainly imported from the Unity asset store and use ProBuild to make some customizations. The scenes take a 1: 1 ratio, and the purpose is to simulate a real world as much as possible. The camera settings also use a first-person perspective to deepen the user's immersion.

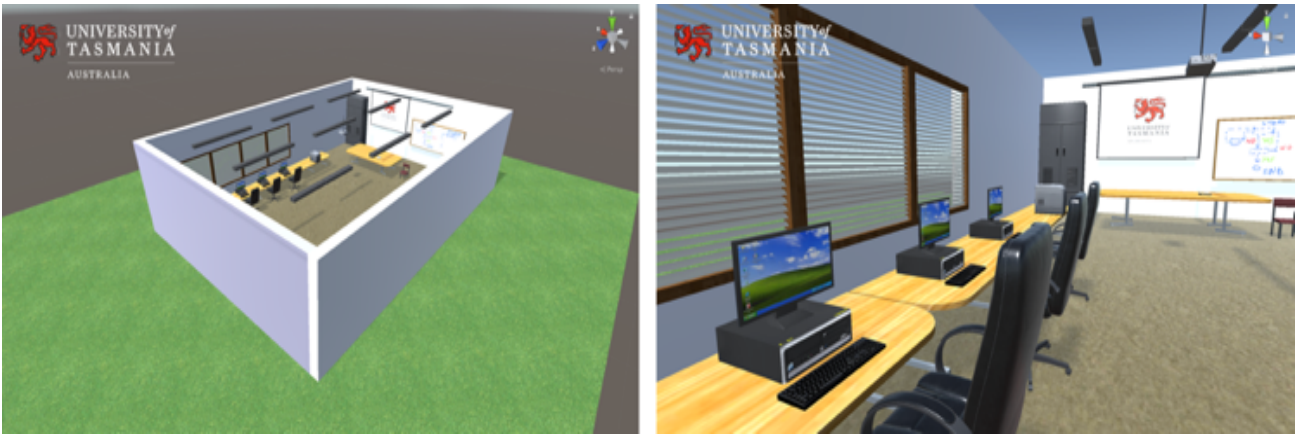

**Figure 5** Model 1 Common Office Room and Interior Settings

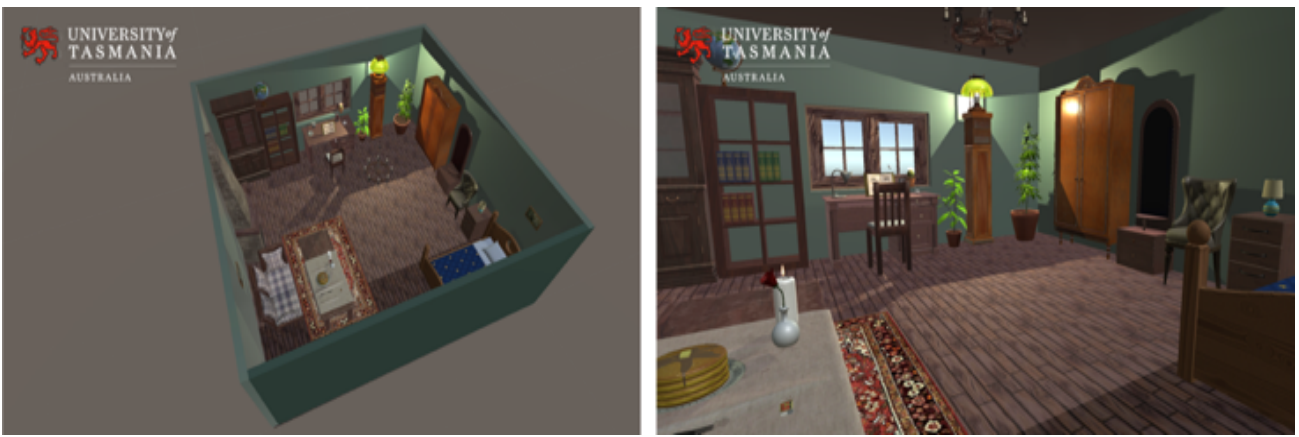

**Figure 6** Model 2 Retro Interior Cottage

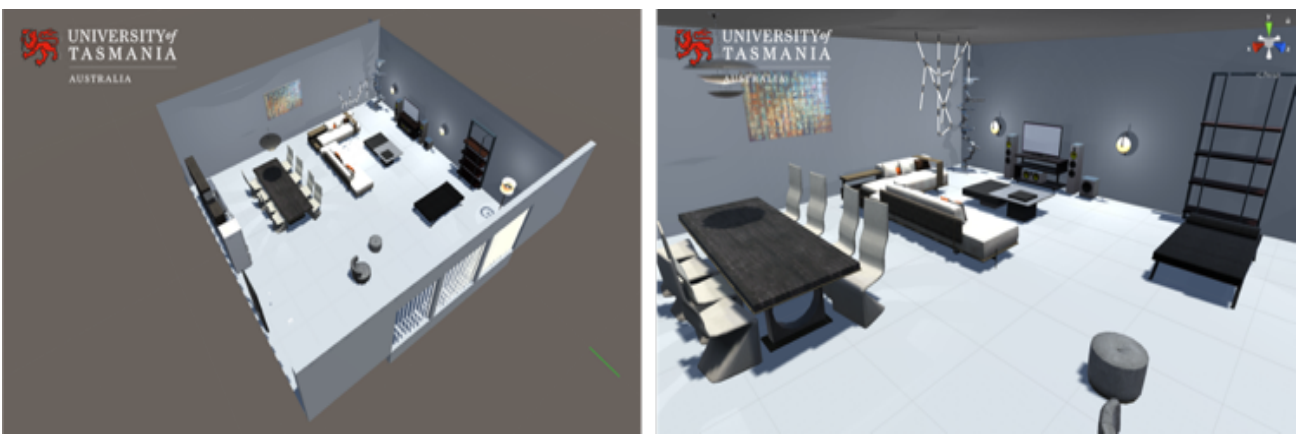

**Figure 7** Model 3 Modern Interior

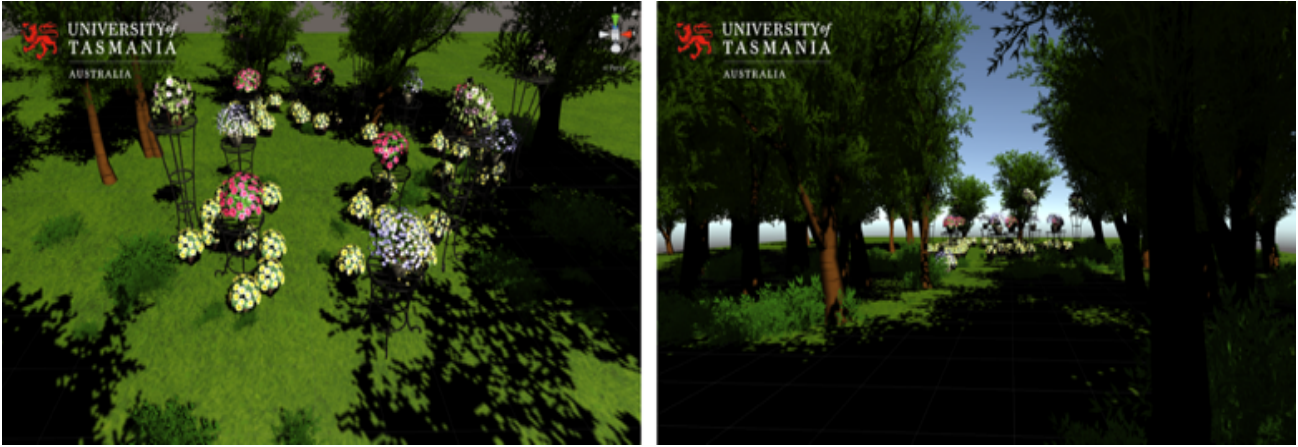

**Figure 8** Modern 4 Outdoor Park

**Step 2. Interacting with VR Equipment**

After developing some VR 3D scenes, next step is to achieve the function of interaction between users and the VR equipment. In this prototype, users can perform a series of basic interactions with the help of the Oculus Go controller. Firstly, when the users wear a VR headset, they can enter a 360-degree 3D scene. Oculus Go can track the user's head in three degrees, as aforementioned in Section 3.1. Therefore, the users can intuitively feel a three-dimensional space by moving the head. Secondly, the users can activate the controller by holding down the controller's front button. After pressing and holding for 3 seconds, the user can select the virtual button set in the scene and trigger some events. For example, users can press and hold to drag, rotate, and throw specific objects. Thirdly, the user can touch the pad on the controller to control the forward, backward, left and right movement. These main interaction modes are also presented in Figure 9.

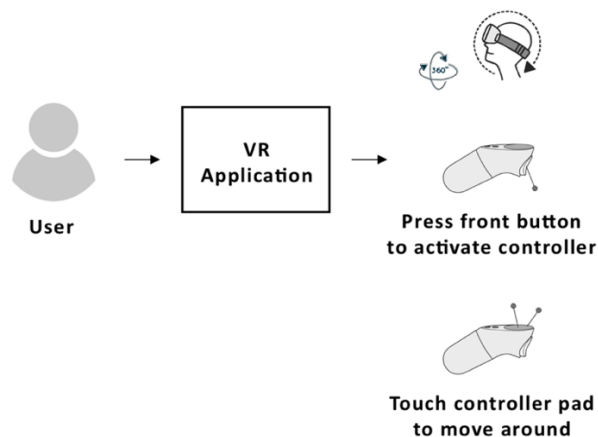

**Figure 9** Main Interaction Modes

Moreover, Figure 10 shows nine specific interactions with the VR scenarios, such as long press the front button of the controller to change scenes or enable a pop-up menu. With the help of these interactions, the capability of users' navigation skills is likely to be assessed and evaluated, which plays a significant role for the following research project about early detection of dementia.

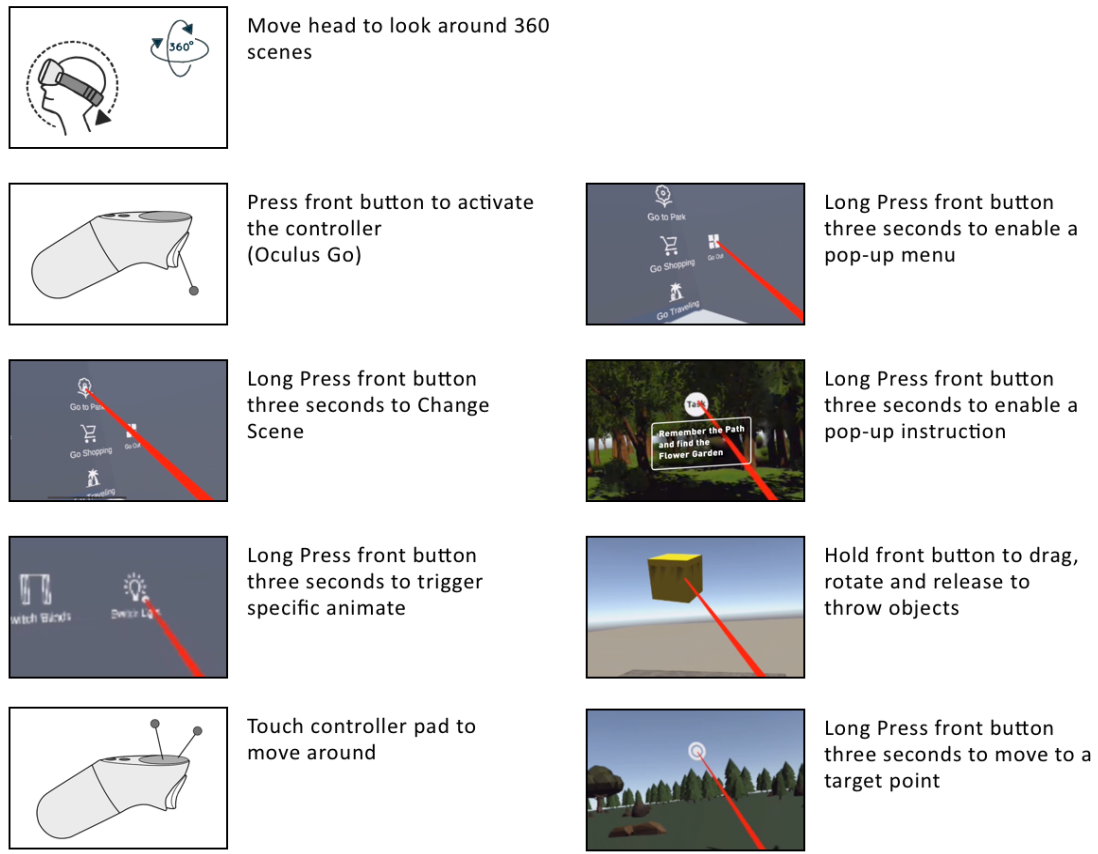

**Figure 10** Specific interactions

## Step 3. Monitoring the interactions of users

In order to assess the performance of the users, a process of monitoring the interaction is necessary, which has the potential to help general practitioners and specialists to diagnosis dementia. This prototype justifies the technical feasibility.

As shown in Figure 11, observers can connect to the Oculus headset via an external device such as a desktop computer or a tablet. This allows the observer to monitor the user's activities in the virtual space and obtain real-time data such as the time the user completed the task and the specific difficulties they encounter. Meanwhile, Oculus's dashboard backend can collect application usage data, such as the time the user spends at each level. Therefore, the observer can evaluate the user's performance in a specific scene. This will expand the evaluation method set and enrich quantitative data, while also avoiding certain risks. For example, subjects can be tested in highly simulated virtual scenarios without accomplishing similar tasks in real-world scenarios. For patients with a severe dementia, this is a safe and reliable method.

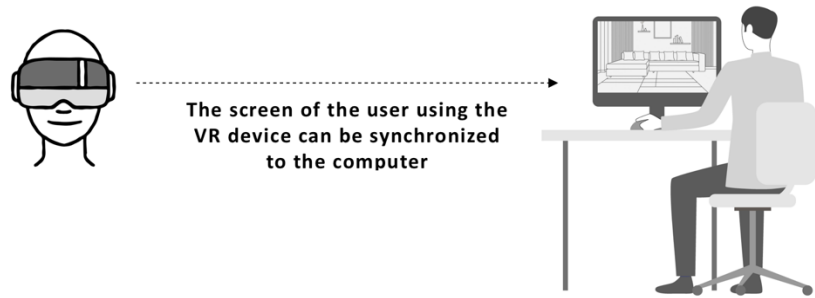

**Figure 11** Monitoring the Interaction between the User and the VR Device

## 4. Contribution

This exploration mainly attempts to apply VR technology for further dementia research project. In addition, a highly detailed record was provided by this technical report, in order to outline a basic technical model. Three primary contributions of this exploration are listed below:

- It was demonstrated that VR technology was feasible to be applied to assess and evaluate the performance of participants.

- It was demonstrated that VR technology has the potential to be utilized for early detection of dementia, which could also avoid the security risks of testing in real-life scenarios.

- It was demonstrated that VR technology was likely to get more participants from culturally and linguistically diverse backgrounds involved in the dementia research.

## 5. Future work

Since this project only focuses on the basic exploration from the technical perspective, the subsequent research project will be conducted under the guidance of specialists of neuroscience, with involvement of participants. Moreover, all models and interactions should be tested rigorously, additionally, thorough evaluation on the models and interactions should be accomplished, in order to ensure that it can play a positive role in dementia diagnosis.